Article

# Mist and Edge Computing Cyber-Physical Human-Centered Systems for Industry 5.0: A Cost-Effective IoT Thermal Imaging Safety System

Paula Fraga-Lamas [1,2,*] , Daniel Barros [3] , Sérgio Ivan Lopes [3,4,5] and Tiago M. Fernández-Caramés [1,2]

1 Department of Computer Engineering, Faculty of Computer Science, Universidade da Coruña, 15071 A Coruña, Spain
2 Centro de Investigación CITIC, Universidade da Coruña, 15071 A Coruña, Spain
3 ADiT-Lab, Instituto Politécnico de Viana do Castelo, 4900-348 Viana do Castelo, Portugal
4 CiTin—Centro de Interface Tecnológico Industrial, 4970-786 Arcos de Valdevez, Portugal
5 IT—Instituto de Telecomunicações, Campus Universitário de Santiago, 3810-193 Aveiro, Portugal
* Correspondence: paula.fraga@udc.es; Tel.: +34-981-167-000

**Abstract:** While many companies worldwide are still striving to adjust to Industry 4.0 principles, the transition to Industry 5.0 is already underway. Under such a paradigm, Cyber-Physical Human-centered Systems (CPHSs) have emerged to leverage operator capabilities in order to meet the goals of complex manufacturing systems towards human-centricity, resilience and sustainability. This article first describes the essential concepts for the development of Industry 5.0 CPHSs and then analyzes the latest CPHSs, identifying their main design requirements and key implementation components. Moreover, the major challenges for the development of such CPHSs are outlined. Next, to illustrate the previously described concepts, a real-world Industry 5.0 CPHS is presented. Such a CPHS enables increased operator safety and operation tracking in manufacturing processes that rely on collaborative robots and heavy machinery. Specifically, the proposed use case consists of a workshop where a smarter use of resources is required, and human proximity detection determines when machinery should be working or not in order to avoid incidents or accidents involving such machinery. The proposed CPHS makes use of a hybrid edge computing architecture with smart mist computing nodes that processes thermal images and reacts to prevent industrial safety issues. The performed experiments show that, in the selected real-world scenario, the developed CPHS algorithms are able to detect human presence with low-power devices (with a Raspberry Pi 3B) in a fast and accurate way (in less than 10 ms with a 97.04% accuracy), thus being an effective solution (e.g., a good trade-off between cost, accuracy, resilience and computational efficiency) that can be integrated into many Industry 5.0 applications. Finally, this article provides specific guidelines that will help future developers and managers to overcome the challenges that will arise when deploying the next generation of CPHSs for smart and sustainable manufacturing.

## 1. Introduction

In Europe, since 2011, the leading paradigm for industrial change was Industry 4.0, which fostered the digital transformation of factories to optimize the processes and make more efficient use of the available resources [1]. However, Industry 4.0 is explicitly focused on industrial improvement, forgetting societal goals and inadvertently ignoring the human cost in terms of human empowerment, employment and social inequality resulting from such process optimization.

When the impact of Industry 4.0 enabling technologies becomes evident in a few years' time, this will be the main issue that will emerge. Consequently, there is a strong need to boost productivity without removing human workers from manufacturing while ensuring their safety and well-being. For such a reason, industry and academia, in part inspired by the Japanese concept of Society 5.0 [2–4], have started to define the principles of the Industry 5.0 paradigm [5–7].

It is important to note that Industry 5.0 is an open and evolving concept, so there are different available definitions on the literature. This article takes as a reference the Industry 5.0 foundations set by the European Commission in January 2021 [8], which have in mind the future of the European industry and the purpose of balancing economic growth and sustainability. In any case, Industry 5.0 is gaining more and more importance worldwide in light of recent events such as the long-standing COVID-19 pandemic, climate change, and the current conflict between Ukraine and Russia. Thus, Industry 5.0 aims to address the limitations of Industry 4.0 while considering current societal trends, which were overlooked by such a paradigm, like social justice or many of the UN Sustainable Development Goals (SDGs) [9].

Industry 5.0 is mostly enabled by the combination of enabling Industry 4.0 technologies like blockchain [10,11], 5G/6G [12], UAVs [13,14], fog and edge computing [15–17], Artificial Intelligence (AI) [18,19], Augmented/Mixed/Virtual Reality (AR/MR/VR) [20], Internet of Things (IoT) and its variants (e.g., Industrial IoT [21], Green IoT (G-IoT)) [22–26], digital twins [27–29] or advancements in fields like energy efficiency or smart materials [5]. Although some Industry 4.0 applications have certain human-centric aspects in mind (e.g., operator safety), in Industry 5.0, the design and application of such technologies goes further than just their mere industrial exploitation, focusing on boosting simultaneously three principles: human-centricity, sustainability, and resilience [30]. Recent literature [31,32] provides reference models and methodologies that detail how Industry 5.0 developments should be managed.

According to the European Commission [8], rather than asking what can be done with such new technologies, we should ask what such technologies can do for the workers. Thus, the Industry 5.0 technologies used should be adapted to meet the diverse needs of industry workers, alerting them and their general practitioners about critical health conditions [3]. To achieve such a vision, Industry 5.0 has to combine humans and machines to boost operational and resource efficiency by combining workflows with intelligent systems in a sustainable manner. The primary focus of Industry 4.0 is intelligent decision-making and connectivity, whereas the goal of Industry 5.0 will be a harmonious coexistence of human-centricity, resilience and sustainability through semi-autonomous manufacturing with humans in the loop. The next generation of robots, commonly referred to as collaborative robots (cobots), will be companions, not simply programmable tools that can perform repetitive tasks. Such robots must be equipped with human-like decision-making mechanisms, which requires sophisticated sensing, localization, and cognition capabilities, as well as increased computing power in embedded platforms [3].

It is important to note that, according to the European Commission [8], Industry 5.0 is not a chronological continuation of the Industry 4.0 paradigm, but a concept that fuses European industrial and societal trends. Thus, Industry 5.0 can be regarded as a complement to the essential characteristics of Industry 4.0 that involves relevant social aspects like social fairness or environmental impact. Therefore, such enhancements allow for refocusing Industry 4.0 principles to create a human-centered and environmentally conscious future.

The previous features can be implemented with the help of IoT/IIoT devices, which currently mainly rely on cloud computing-based platforms that perform centralized data processing. Nevertheless, such a centralized implementation has several disadvantages (e.g., it is not inherently energy efficient, it usually has a relatively high delay, and it may be disrupted by attacks, vulnerabilities, maintenance activities, or a potential overload of the cloud). As a result, novel decentralized and distributed computing paradigms have recently arisen. One of them is edge computing, which is able to offload the cloud from performing tasks that could be performed by computing devices located at the network edge [33]. In that way, latency response together with the number of packets exchanged with the cloud are substantially reduced. Mist computing goes a step further, moving part of the data processing from devices on the edge to the end IoT/IIoT devices [34]. This is possible thanks to the progress in recent years of embedded computational power, which has allowed for the creation of more powerful and efficient IoT/IIoT end nodes whose capabilities avoid transmitting data to the higher layers (e.g., to edge devices or to a remote cloud).

The concept of a Cyber-Physical System (CPS) is also relevant for Industry 5.0 factories, which is a system built from the integration of embedded computing devices, physical processes, networking, and computation deployed through a communication infrastructure. As a result, CPSs need to rely on most of the aforementioned technologies (e.g., G-IoT, edge computing, or Edge-AI). CPSs are extensively used in a variety of systems, application domains and environments [35,36] in the context of smart manufacturing, and, increasingly, they have to be able to proactively and autonomously adapt while in use (e.g., to reason, to self-learn, to cooperate) to the changing needs of industrial operators (e.g., due to mobility), to unanticipated situations, or to new environmental conditions [37].

Currently, CPS's full autonomy has not been achieved due to its high cost and complexity, and may even not be desirable under the Industry 5.0 paradigm. In such a paradigm, the role of humans is essential to collaborate, to supervise, and to be part of the feedback control loop. The best approach to CPSs is that humans and machinery/robots work collaboratively in a semi-automated way, thus combining the strength, accuracy, consistency, and speed of machinery with the adaptability, ingenuity, judgment, creativity, and dexterity of human workers. In the literature, there are several terms that have been used to refer to such systems: cyber-human systems [38], Human-in-the-Loop CPSs (HiLCPS) [39–41], Human-centered CPSs, Human Cyber-Physical Systems (HCPSs) [42,43], Cyber Physical Human Systems (CPHSs) [44,45], or even the term Cyber-Physical-Social Systems [46–49], which considers social networks as part of the different user information sources. In the rest of this article, we will use CPHS to refer to CPSs that leverage people's capabilities to meet the goal of advanced complex industrial systems by ensuring contextual awareness, higher efficiency, enhanced performance, high interoperability, self-cognition, and human error reduction. The main attributes of the so-called Industry 5.0 CPHS are illustrated in Figure 1. The interested reader can find in [42] a detailed review of the implications of CPHSs in the evolution of manufacturing systems.

This article, after analyzing the characteristics and the current state of the art of CPHSs, details the development of one of such systems. The proposed CPHS enables a holistic awareness of the overall factory floor, thus increasing both safety and manufacturing predictability and adaptability to new situations over time. It makes use of a hybrid edge computing architecture with smart mist computing nodes with low-cost thermal imaging sensors.

Specifically, the contributions of this article are the following:

- The essential concepts for the development of CPHS systems are detailed.
- Some of the most relevant works on CPHSs have been analyzed. Their main design requirements and key implementation components have been identified.
- The major challenges for the development of Industry 5.0 CPHSs are outlined.
- An Industry 5.0 CPHS use case designed to increase industrial operator safety in human-machine collaborative environments is described in detail and its performance is evaluated. The use case makes use of a hybrid edge computing architecture and low-cost thermal imaging sensors to provide an adequate trade-off between cost, accuracy, resiliency and computational efficiency. To the best of our knowledge, no similar CPHS has been previously described in the literature.
- Key findings, challenges and relevant guidelines are outlined in order to help future Industry 5.0 researchers and engineers.

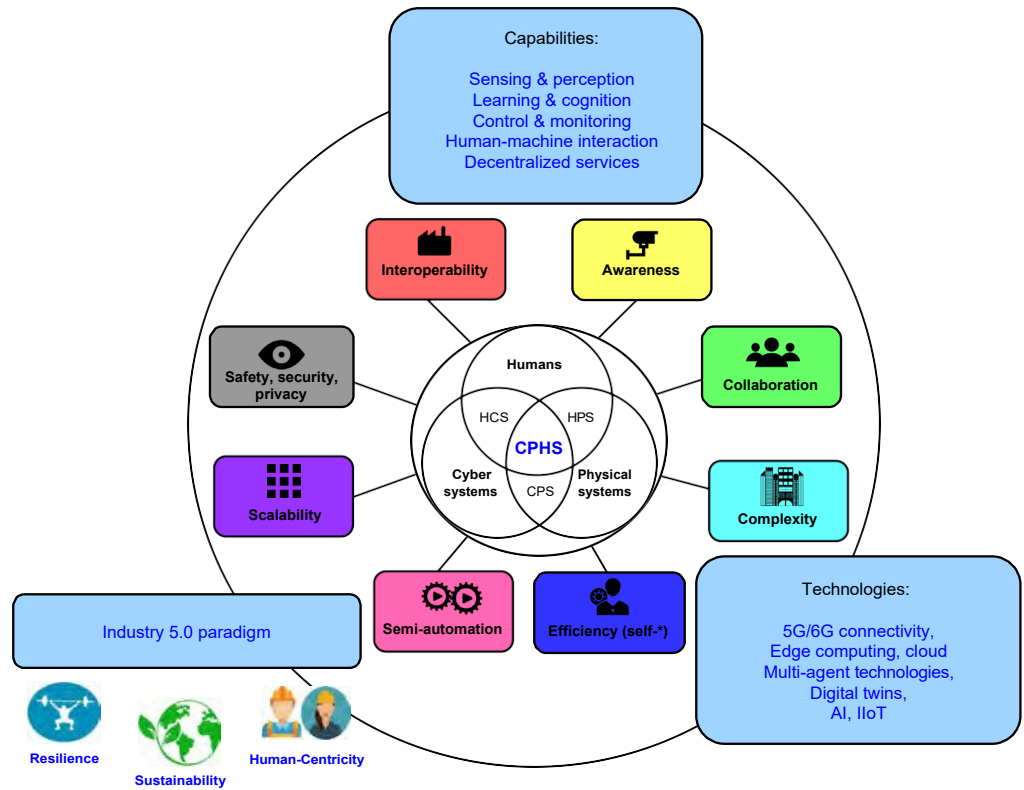

**Figure 1.** Main attributes of an Industry 5.0 CPHS.

The remainder of this article is structured as follows. Section 2 provides a thorough analysis of the state of the art on human-centric CPSs and on their degree of compliance with the Industry 5.0 paradigm. In addition, the most relevant human-detection systems for industrial environments are described. Section 3 describes the structure and main layers of a modern CPHS communications architecture. Section 4 details the design, implementation, and evaluation of the selected application case: a cost-effective edge/mist computing system based on the use of IoT thermal imaging nodes for increasing safety in Industry 5.0 scenarios. Section 5 presents a discussion on the lessons learned from the experiments, outlines the main challenges, and proposes future work directions. Finally, Section 6 is devoted to the conclusions.

## 2. Related Works

### 2.1. Towards Industry 5.0 Cyber-Physical Systems

In recent years, the research on CPSs has become increasingly popular in both academia and industry, but, unfortunately, just a few works present CPSs specifically designed for meeting the requirements of Industry 4.0/5.0. One such work is described by Thakur et al. [50], who analyzes some critical issues related to CPS design. Other interesting developments are detailed in [51], where the authors propose a framework emphasizing security aspects, and in [52], where a five-level CPS architecture for Industry 4.0 is proposed.

A key issue in CPSs is their enabling Industry 5.0 technologies. Leitao et al. [36] present an overview of technologies such as Multi-Agent Systems (MAS), Service-Oriented Architecture (SOA), and cloud-based systems. This article also outlines the complexity of developing CPSs with humans-in-the-loop. In the case of [53], the authors focus on Open IoT, where devices are shared by various services in the network and the CPS should be constructed automatically through the cooperation of multiple devices.

Despite the fact that CPSs are very helpful for Industry 5.0, their widespread adoption has been hindered by the mismatch between the characteristics of physical processes and their abstraction [54]. In addition, there are still many open research challenges [55–57] like sustainability, complexity, human–robot collaboration, fault tolerance, data security, interoperability, and flexible design and operation. Cyber-resilience is also essential for Industry 5.0, and therefore, different authors have studied it. For instance, Cyber-Physical Energy Systems (CPES) and their cybersecurity are thoroughly analyzed in [58]. In addition, another challenge is how to process data for decision-making. Such an issue has been addressed by Cao et al. [59] through the integration of CPSs with edge computing. They have analyzed current works on edge computing-assisted CPSs; and introduced additional challenges regarding Quality of Service (QoS) optimization, energy consumption, and security.

*2.2. Cyber-Physical Human Systems*

In the research community, there is an increasing effort to understand relevant aspects of the human factor (e.g., performance, risks), to be able to model and predict human behavior [60]. As it is illustrated in Figure 2, the role of humans in the context of CPHS challenges traditional CPSs in the following aspects [44]:

- Knowledge: Human performance and behavior is affected by factors like available information and training, but also by different factors like stress, mood, fatigue, or motivation.
- Perception: The process of data collection can be performed by hardware sensors and/or humans (i.e., social sensors).
- Cognition: Regarding situational or context awareness, people observe and process information differently than computers.
- Learning: Humans do not perform the same task in exactly the same way every time, therefore, there are different levels of predictability. In addition, humans have high adaptability, so they can easily adapt to changing environments.
- Actuation: Remote monitoring and control to provide safer and human-friendly human–machine interactions.

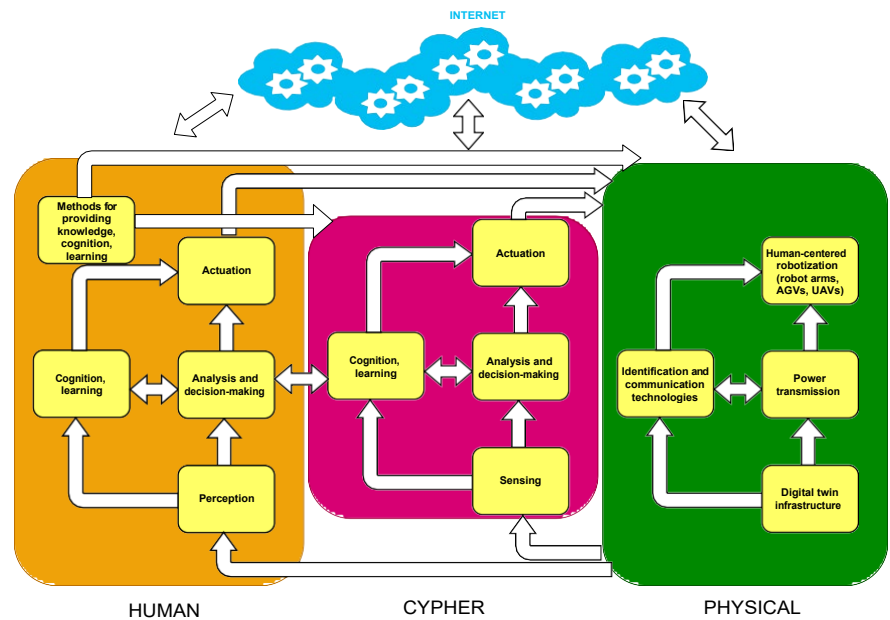

**Figure 2.** Main aspects of CPHS.

As a result, it is very important to collect data about the interaction of humans with machines, devices, or processes to predict probable human behavior under different circumstances. The input data for human factor assessment (i.e., Human Reliability Assessment (HRA) [61]), is mostly obtained from observations during activities, which can be first modeled and then simulated to predict certain behaviors [62]. Other authors go further and propose to develop adaptive CPHSs to improve cognition [63].

Unfortunately, most previous research on CPHSs has not sufficiently considered humans as service providers. For instance, in [64], the authors propose a conceptual SOA ontology model for making use of humans as a service provider for CPS. In such a paper, the accuracy, completeness, flexibility, clarity and consistency of the ontological model are assessed. Additionally, the service composition model is evaluated in a medical use case.

Some authors remark that the development of CPHSs is very complex, since the integration between humans and machines in an efficient and intuitive manner is not easy [62]. Authors like Broo et al. [55] underline the importance of transdisciplinary and introduce three mindsets as a holistic approach: system mindsets (i.e., interconnectedness, feedback, adaptive capacity/resilience, emergence and self-organization), future thinking and design thinking mindset (i.e., devise, observe, collaborate, learn, prototype, test, operate, integrate).

Regarding frameworks and reference models, Razaque et al. [65] propose a framework of a human-centric mobility-aware CPS model for healthcare. In addition, such an article reviews the main related works that deal with CPS security. In addition, Ma et al. [66] present a CPHS reference model and discuss three specific aspects: semantics, interaction, and iteration (e.g., remembering prior knowledge to handle uncertainty and resolve ambiguity). Moreover, they address major challenges from the standpoint of data characteristics (e.g., real-time, high dimension, and frequency). Furthermore, the progress of the state of the art is examined in relation to three industrial sectors: space, transportation, and smart grids. As a result of such a work, the authors conclude that most of the existing frameworks have not considered most of the necessary features of a CPHS and effectively handle the challenges of the collected data.

Other relevant works are presented in [67,68]. In the case of [67] the authors propose a framework that enables the integration of digital twins. Regarding the work detailed in [68], it is focused on the development of smart factories, presenting a framework for a smart shop floor that includes three primary models: one of the smart machine agents, a self-organizing model responsible for optimally matching resources with tasks, and a self-adaptive model, which aims to monitor processes and autonomous manage disruptions.

With respect to the intelligent capabilities of CPHSs, most of the current solutions rely on compute-intense algorithms. For instance, Wang et al. [69] focus on learning capabilities, mainly proposing solutions that involve AI algorithms. In the case of Ren et al. [70], an adaptive ML-based CPHS with a 12.5% detection accuracy is proposed. Such a proposal leverages the causality of human–machine interactions and is initialized by a priori knowledge and pre-processed public datasets.

Turning physical assets into CPHS through digital twins is also a hot topic. For example, Broo et al. [71] recently presented a review of a digital twin architecture together with a case study implementation of an infrastructure asset.

*2.3. New Communications Architectures for CPHSs and CPSs*

As previously mentioned, novel computing paradigms have recently emerged in order to face the disadvantages of the cloud computing platforms traditionally used by CPS and CPHS. One such paradigm is fog computing, which uses low-power devices on the edge [72]. In the case of needing to perform complex processing on the edge, cloudlets can be used [73]. Mist computing goes a step further and offloads computation directly to IoT/IIoT end nodes, which are self-aware and able to communicate and coordinate with each other to perform complex tasks collaboratively [74].

Decentralization is also an interesting feature to improve the resilience of Industry 5.0 CPHSs. Currently, most CPSs make use of a central component (often a service that is executed on the cloud) that manages coordination tasks and controls access to external applications and remote-user interfaces. However, decentralized architectures, in which every node plays its part in line with a decentralized protocol, are emerging [75,76]. Such architectures are suitable for networks that are jointly deployed by several entities (such as businesses, organizations, governments, or individuals), and there is no need to rely on a single coordinating entity.

Finally, it is worth mentioning that the previously mentioned types of architectures are not mutually exclusive, so, depending on the particular application, hybrid architectures can be built to provide the desired functionality. For instance, the system proposed in this article mixes edge and mist computing, thus relying on edge devices for handling time-sensitive intensive computing tasks, and using mist nodes to collect and process data in a collaborative self-aware autonomous manner.

### 2.4. Human Detection and Safety in Industrial Scenarios

The need to recognize unusual objects in certain areas has driven the development of new automated solutions to identify and notify potential security threats as quickly as possible [77,78]. In particular, most of the recent research focuses on monitoring techniques for safety in human–robot collaboration [79].

Different technologies have been previously used to detect human presence and thus protect operators from the hazards that may occur due to working in industrial scenarios. For instance, in [80], the authors propose a safety system for smart factories that makes use of eight-megapixel cameras and ultrasound sensors to detect humans. For such a purpose, the authors compare the performance of different identification algorithms (e.g., HoG (Histograms of Oriented Gradients), YOLOv3, Viola James), evaluating their latency when using different resolutions.

In the case of [81], a survey is presented on the currently available thermal sensors that are commonly used in different industries. Specifically, the paper discusses the use of thermal sensors in various manufacturing industries, such as the automotive and manufacturing industries, as well as in Heating, Ventilation, and Air Conditioning (HVAC) systems. In addition, the authors compare the performance of different detection algorithms and techniques (e.g., AdaBoost, GNB (Gaussian Naive Bayes), KNN (K-Nearest Neighbours), and SVM (Support Vector Machine)), and conclude that, in the selected experimental setup, they are able to achieve a 100% detection accuracy when mounted on the ceiling and 98.6% when mounted on a wall.

Another interesting work on the use of thermal imaging sensors for occupancy estimation is presented in [82], where the authors compare three thermal imaging sensors (MLX90393, Grid Eye and FLIR Lepton 2.0) in terms of their cost, resolution, field of view, operation voltage or refresh rate, among others. The authors also propose a solution that makes use of a combination of active frame analysis, component analysis, feature extraction, and classification that can be implemented in various smart building applications. Similarly, in [83], Braga et al. present the assessment procedures and results of comparing the performance of two low-cost thermal imaging cameras (MLX90640 and FLIR Lepton 2.5) with a more expensive flagship medical screening device. The obtained results show that, although thermal imaging sensors are far from achieving the performance of medical-grade devices, they can be used for the detection of elevated temperature events in humans.

Finally, it is worth mentioning the work described in [84], where a multi-modal sensor array for safe human–robot interaction and mapping is presented. The sensor array consists of ST Micro-electronic VL6180X Time-of-Flight (ToF) sensors (which measure the time the light needs to travel to an object and back to the sensor) and Triaxis magnetic-field sensors (MLX90393) for force sensing and localized contact detection.

### 2.5. Analysis of the State of the Art

The authors of [31] conducted a very interesting literature review and identified a number of relevant characteristics such as real-time communications, intelligent automation, operational and resource efficiency, system integration and interoperability, among others, through which Industry 5.0 can promote the values of sustainable development. These functions can be categorized in relation to human-centricity, sustainability and resilience, and are taken as the basis for the follow-up analysis of the state of the art.

Table 1 compares the main Industry 5.0 characteristics of the most relevant analyzed CPSs and CPHSs. As it can be observed, the majority of the documented solutions are aimed at implementing smart manufacturing systems [36,52,56,57,68,70], and there are also a relevant number of papers that describe CPSs/CPHSs for the healthcare industry [51,64,65], for smart grids and smart energy production systems [50,54,58–60,66], and for smart infrastructure monitoring and control [55,67].

Moreover, there is also a relevant number of papers that propose developments oriented towards providing human-centricity to current CPSs [44,62,63,69]. However, only one of the analyzed developments [65] provides human-centricity together with the other two main principles of the Industry 5.0 paradigm (sustainability and resilience). The rest of the analyzed works only comply with one or two of the mentioned principles and, in some cases, the principles are only fulfilled partially (for instance, in [36,55,58]).

*2.6. Contributions and Novelty of the Proposed System*

In contrast to the state of the art, the CPHS presented in this article has been specifically devised for being compliant with the three Industry 5.0 principles:

- Human-centricity: The whole system has been specifically designed for preserving industrial operator safety in human–machine collaborative scenarios. Thus, specific human-detection algorithms have been developed to determine when dangerous situations may happen in industrial environments in order to control certain IIoT devices (e.g., cobots in operation).
- Sustainability: The devised system was conceived to be implemented by using low-cost and low-power devices. In addition, the proposed solution is based on a hybrid edge/mist communications architecture that seeks to minimize data exchanges with the remote cloud servers, thus decreasing their energy consumption and the power dedicated to communications. Furthermore, since the developed human-detection algorithms do not make use of machine learning or other Artificial Intelligence techniques, the usually energy-consuming training stage and the required hardware (e.g., computers with power-hungry GPUs) are not necessary.
- Resilience: The previously mentioned edge/mist computing architecture provides IIoT node redundancy, and it allows for distributing the computational load, thus decreasing the potential saturation of the cloud. In addition, the architecture avoids unnecessary data exchanges with the cloud, so it prevents attacks on the system communications (e.g., man-in-the-middle attacks). Furthermore, since the proposed solution uses thermal imaging sensors instead of regular webcams, it deals with potential privacy issues, which have been a traditional cybersecurity concern for many industrial surveillance systems.

As a result of the previously described design decisions, the proposed solution provides an adequate trade-off between cost, accuracy and computational efficiency that can be easily deployed into IoT architectures, thus being able to operate in a wide variety of Industry 5.0 scenarios. To the best of our knowledge, no similar CPHS has been previously described in the literature.

Table 1. Main characteristics of the most relevant CPS/CPHSs.

| Reference | Type of System | Application Field | Human-Centricity | Sustainability | Resilience | Other Relevant Features |
|---|---|---|---|---|---|---|
| [50] | CPS | Design automation, System synchronization, Dynamic Voltage and Frequency Scaling | × | | | Discussion on the design of a smart mission-critical CPS. |
| [51] | CPS | Healthcare | × | × | | Framework with a Security Layer with Open Authorization (OAuth), User-Managed Access (UMA), and Self-Sovereign Identity (SSI). |
| [52] | CPS | Manufacturing, Prognostics and | × | × | | Description of the flow of information on a 5-layer CPS architecture. |
| [36] | CPS | Manufacturing, Electronics assembly, Continuous processing, Energy management | Partially (theoretical) | × | | Overview of key aspects of ICPSs, future research, as well as implementation challenges based on the experience of four European innovation projects. |
| [53] | CPS | Experiments with IoT devices: two lighting devices and a web camera. | × | | | CPS for Open IoT. It includes a method of actuator control optimization based on service satisfaction and system operation cost. It allows for automatically constructing the CPS by using non-configured IoT devices or robots connected to the network. |
| [54] | CPS | Smart grid infrastructure, Household energy consumption, Behavior change | | | × | CPS developed with data mining techniques to provide households with feedback messages about energy use. Evaluation of performance of dynamic behavioral reference group classification. |

Table 1. *Cont.*

| Reference | Type of System | Application Field | Human-Centricity | Sustainability | Resilience | Other Relevant Features |
|---|---|---|---|---|---|---|
| [55] | CPS | Digital twin in smart infrastructure (E-type steel half-through railway bridge) | Partially (information collected is presented in a human-centered way, metrics) | × | | Literature review of digital twin architectures. |
| [56] | CPS | Manufacturing | × | × | × | Challenges of CPSs in manufacturing |
| [57] | CPHS | Hazard Manufacturing | Yes (human-robot collaboration) | × | × | Design and testing of a remote robot control system (i.e., from robot to robot) and a model-driven display system (i.e., to ease the understanding of the production context by the operator) that can work in several modes. |
| [58] | CPS | Critical infrastructure, energy systems | Partially (HMI) | × | | Framework for security evaluation (model, simulate, assess and mitigate) illustrated by four attack case studies. Risk assessment analysis. |
| [59] | CPS | - | × | Partially (energy consumption minimization) | | Edge and edge-cloud computing CPS. Survey of the state of the art and analysis of challenges in service latency reduction, Quality of Service (QoS) optimization, energy consumption minimization, trade-off service latency and energy consumption, security and privacy enhancement, and reliability augmentation. Description of future research directions. |

**Table 1.** *Cont.*

| Reference | Type of System | Application Field | Human-Centricity | Sustainability | Resilience | Other Relevant Features |
|---|---|---|---|---|---|---|
| [60] | CPHS | Critical infrastructure, nuclear power plant | | × | | Identification of domain requirements (e.g., data collection, data processing, computational modeling, simulation), challenges, and potential solutions to create a human-centered automation system that supports resilient nuclear power plant outage control. Enhanced performance of handoff monitoring and control between tasks and to respond to contingencies in workflows during the outage. Research roadmap for human-centered automation in construction. |
| [44] | CPHS | - | | × | × | Human service capability description model to help in the integration of people in CPHSs. The model focuses on the structure of data representing a person's capability. |
| [62] | CPHS | - | | × | | Measurements and modeling of human behavior, prediction of human response and his/her dynamic properties. |

Table 1. *Cont.*

| Reference | Type of System | Application Field | Human-Centricity | Sustainability | Resilience | Other Relevant Features |
|---|---|---|---|---|---|---|
| [63] | Adaptive CPHS | - | | × | × | Three types of adaptive CPHS: human controls CPS, CPS monitors human and act when needed, and human monitoring of the CPS that acts as controller. Discussion of key learning and adaptation techniques. Current challenges of adaptive CPHS (e.g., cognitive modeling and machine learning in the control loop). Functional reference architecture to guide the developers of adaptive CPH systems. |
| [64] | CPHS | Healthcare | | × | × | SOA ontology model for CPHS: human characteristics and their dynamics, as a service provider or collaborator with the machine (e.g., sensing, processing, actuating, and promote adaptation for other nodes). Ontology evaluation. |
| [65] | CPHS | Healthcare | | | | Prototype for a privacy-aware secure human-centric mobility-aware (SHM) model to analyze physical and human domains in IoT-based wireless sensor networks (WSNs). |
| [66] | CPHS | Smart space, transportation and grid | | × | × | Reference model and key challenges (semantic, interactive, iterative). Analysis of different aspects of data characteristics. |
| [67] | CPS | Smart infrastructure (bridges, railways and transport systems) | | | × | Systems architecture to integrate different digital twins. |

**Table 1.** *Cont.*

| Reference | Type of System | Application Field | Human-Centricity | Sustainability | Resilience | Other Relevant Features |
|---|---|---|---|---|---|---|
| [68] | CPS | Manufacturing, smart shopfloor | × | | | CPS for intelligent shopfloor with high autonomy, adaptability, and efficiency. Manufacturing machines are modeled as smart agents. Self-organizing capabilities to maximize resource efficiency. Self-adaptive mechanism when exceptions occur. |
| [69] | CPHS | - | Partially | × | × | Review-based and interaction-based learning and intelligence in CPHS. |
| [70] | CPHS | Manufacturing, worker-machine interaction | | × | | Adaptive ML based smart manufacturing interactive CPHS. The ML model self-evolves during deployment with the streaming data in a self-labeling manner. Enhanced accuracy for human machine interaction detection by up to 12.5%. |
| This work | CPHS | Industrial operator safety | | | | Use of a novel edge/mist computing and efficient human-presence detection algorithms that can be executed in low-power IoT nodes. |

## 3. CPHS Communications Architecture

*3.1. Essential Components*

A state-of-the-art industrial CPHS is usually composed of the three layers shown in Figure 3. All the industrial devices (such as IIoT devices and platforms, sensor networks, collaborative robots (COBOTs), Automated Guided Vehicles (AGVs), drones, or Industrial Augmented Reality (IAR), Industrial Mixed Reality (IMR) or Virtual Reality (VR) devices) and heavy machinery make use of edge computing layer services. In the Edge Computing Layer, fog services are provided by Single-Board Computers (SBCs), which are scattered throughout an industrial plant and serve as gateways, while heavy-processing tasks are performed by cloudlets [85]. Finally, at the top layer is the cloud. The next subsections provide more details on the mentioned layers.

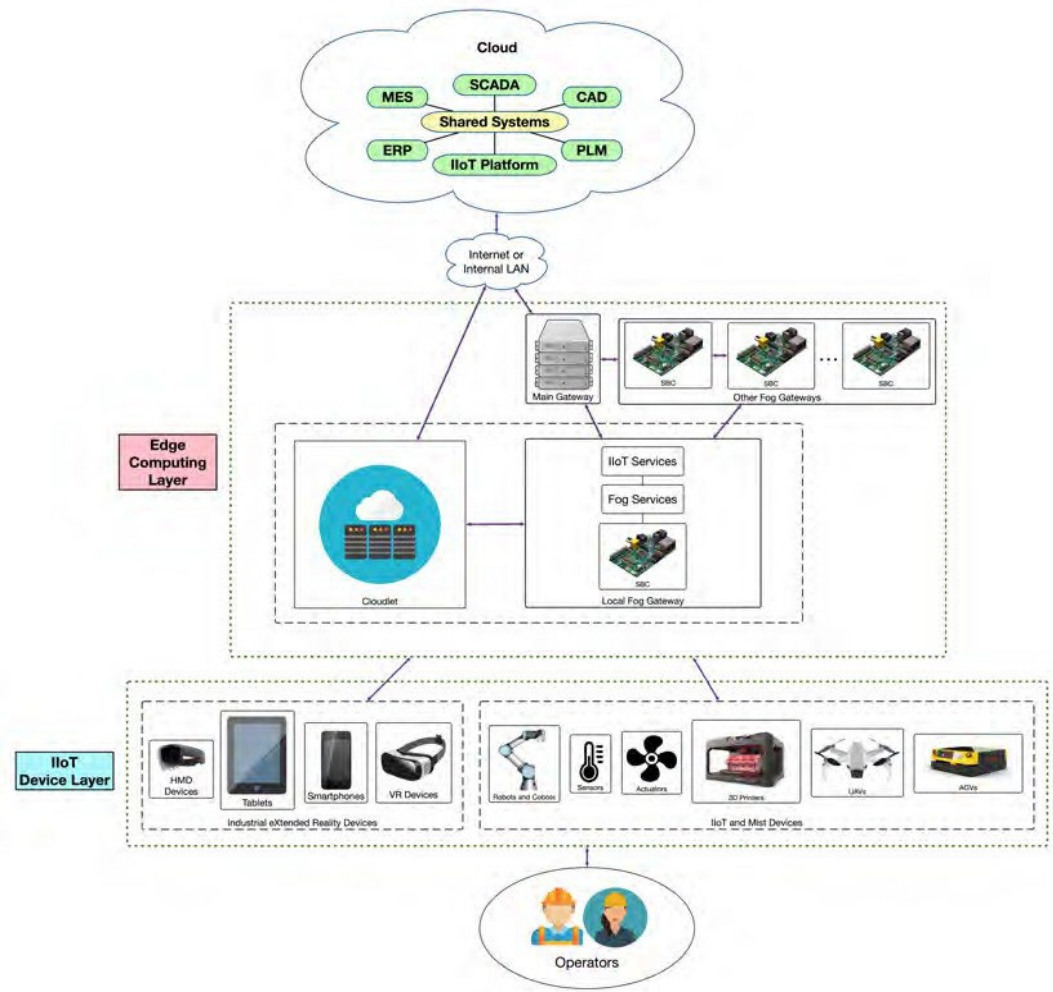

**Figure 3.** Edge Computing based communications architecture for Industry 5.0 applications.

*3.2. IIoT Device Layer*

IIoT devices (e.g., wearables, tracking sensors, biometric sensors, cameras), COBOTs, UAVs, AGVs, IAR/IMR/VR devices, Human–Machine Interfaces (HMI), embedded systems, and transducers constitute the actual IIoT node layer. The data provided by such a layer is monitored, processed, and analyzed by a service-oriented architecture that determines which actions should be performed. In addition, operators can also collect directly complex information (e.g., failures), which may be challenging to be processed directly by physical sensors.

Mist computing nodes are also included in this layer and are distinguished from the rest of the IIoT devices by (1) their capacity for being self-aware; (2) their ability to perform part of the processing traditionally carried out at higher layers; and (3) being able to collaborate autonomously with other mist devices to implement joint complex processing tasks.

*3.3. Edge Computing Layer*

As it was previously mentioned, traditional centralized communications require an external server (usually a remote cloud), which responds to the requests sent by IIoT devices. The problem is that such a way of communicating incurs in delays that are longer than the ones involved when communications are strictly performed inside a local network. Moreover, in a cloud-centric approach, the increase in the number of IIoT nodes might derive into traffic bottlenecks when the cloud receives multiple concurrent requests.

The aforementioned issues can be tackled by using edge computing, which offloads the cloud capabilities to IoT/IIoT devices located at the network edge that provide edge services. As a result, the data processed by the mentioned edge services are generated by the IIoT device layer, as well as by the operators that interact with the CPHS. Such elements will provide different sources of data (i.e., information, knowledge) that will be continuously reported to the fog services (e.g., data generated through web requests).

In addition, the Edge Computing Layer contains cloudlets [86], which are capable of performing computationally demanding tasks like video or image processing. A cloudlet's response latency is significantly lower than when utilizing typical cloud computing systems since it is proximate to the IIoT devices that request its services. Moreover, the use of edge computing allows for performing certain complex tasks (e.g., perception of the surrounding environment and analysis of the data to undertake reasoning and decision-making) without the need for delegating them to higher layers. In addition, the architecture shown in Figure 3 facilitates scalability, so it is straightforward to add new nodes and services to the CPHS as they grow with time.

By significantly reducing the amount of hardware required to deploy the architecture, edge computing systems result in lower costs and less energy consumption. Additionally, latency is decreased by reducing the number of network nodes that the requests have to go through to reach the end IoT devices. Furthermore, the involvement of the Edge Computing Layer provides a high degree of autonomy for data processing and decision-making, enabling it to operate via alternative paths if some nodes do not work properly.

*3.4. Cloud*

An industrial company cloud is located at the top of the architecture, as it can be observed in Figure 3. The cloud is usually responsible for running and managing compute-intensive services and systems like the ones related to industrial auto-identification, Real-Time Locating Systems (RTLS), Manufacturing Execution Systems (MES), Supervisory Control And Data Acquisition (SCADA), Computer-aided design (CAD) software, Enterprise Resource Planning (ERP) software, IIoT platforms, Product Lifecycle Management (PLM) software, or digital twins.

**4. Application Case: CPHS-Based Thermal Imaging Sensors for a Smart Workshop**

In order to illustrate the benefits of CPHSs with a practical example, this section describes the development of a cost-effective edge/mist computing safety system for an Industry 5.0 environment where humans and collaborative robots must interact and work together safely. The proposed low-cost system has been designed to be deployed in factory floors for detecting and counting humans in distinct areas of an image in real time. Thermal imaging is used rather than conventional RGB cameras to increase the chances of detecting humans in situations with low visibility [87]. In addition, the system sensitivity to non-medical grade human-related activities is increased due to focusing exclusively on detecting heat sources [83]. Specifically, as it will be described in the next subsections, the developed system makes use of a hybrid method to determine if: (1) a human is present in a thermal image; (2) a human is crossing a virtual fence; or (3) a human is located in a specific area of the image, that may be associated with a restricted physical area.

## 4.1. General Overview of the Proposed System

The proposed CPHS is a cost-effective solution capable of being installed in factory floors, offices, medical facilities and even outdoors. In addition, the devised solution is also intended to be a one-size-fits-all type of solution able to operate in different environmental conditions with minimal intervention and adjustments. Regarding the capabilities of the proposed solution, the main focus of its design was to monitor specific areas of an image (as it is shown in Figure 4), with the purpose of implementing a virtual barrier or any other similar solution, as well as allowing for estimating the occupancy of a room accurately. For these purposes, a combination of image processing methods was devised and implemented, as it will be later described in Section 4.3.

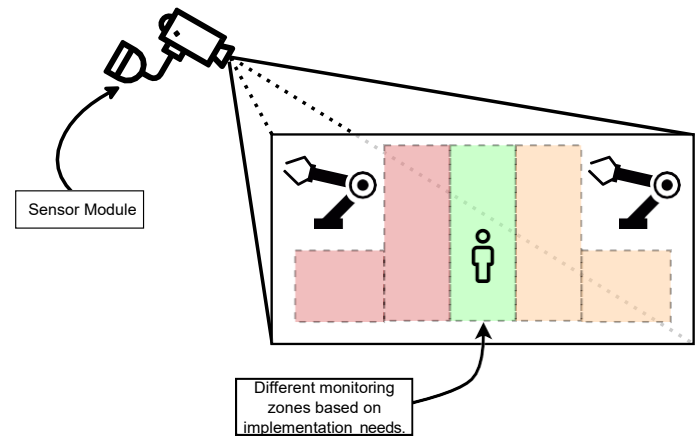

**Figure 4.** Example of factory floor zoning based on the definition of zones surrounding operating collaborative robotic arms, and safe zones where people can occupy in their daily tasks.

## 4.2. Implemented Communications Architecture

A specific communications architecture was devised for the selected industrial use case, based on the architecture previously described in Section 3. Such an architecture is shown in Figure 5. As it can be observed, there are three main layers:

- IIoT Device Layer: It is composed of IIoT sensor modules and robotic arms controllers, which act as mist computing devices that collaborate to prevent industrial operators from being exposed to dangerous situations when working next to robotic arms. These mist computing devices can avoid exchanging image data over the network with edge computing devices or with the cloud, thus benefiting from:
  - Low latency: Since most of the processing is performed locally, a mist computing device can respond faster.
  - Improved local communications with other IIoT nodes: Mist devices can implement additional logic to communicate directly and autonomously with other IIoT devices and machines. Since there is no need for intermediary devices (e.g., servers or the cloud), responses and data exchanges are faster and less traffic is generated to the higher layers.
  - Fewer communication problems: Less interference in complex industrial environments, given that local processing prevents continuous and persistent communications with local edge devices, or remote clouds [88].
  - Fewer privacy issues: There is no need to send camera images to devices over the network, so possible attacks on those devices or man-in-the-middle attacks can be avoided and thus prevent other image leaks.
- Edge Computing Layer: it is able to provide fast-response services to the IIoT/mist devices. Nonetheless, since the development presented in this article is focused on analyzing the performance of the created mist computing devices, no specific data processing services were deployed in this layer.
- Cloud: It behaves as in traditional edge computing-based architectures, thus handling requests that cannot be dealt with locally.

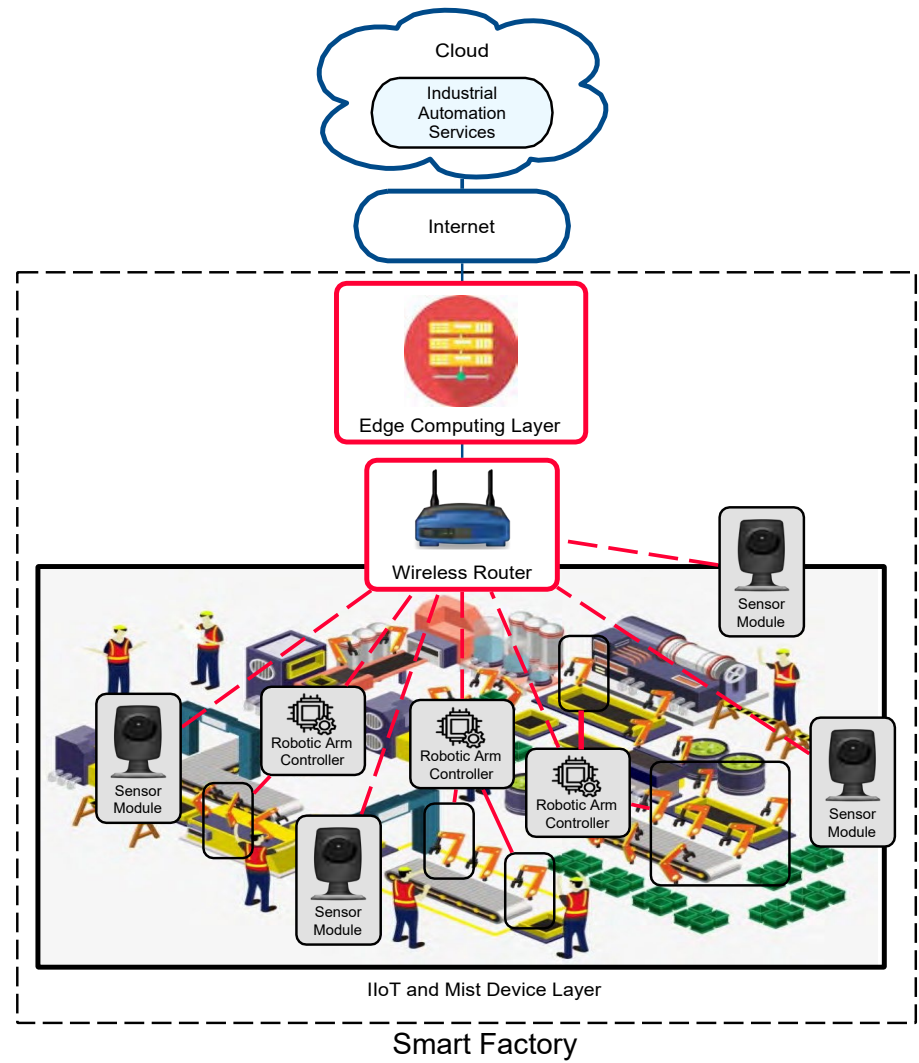

**Figure 5.** Communications architecture of the implemented CPHS.

*4.3. Implementation*

4.3.1. Mist Node Hardware

Three different hardware have been selected for implementing each mist computing node of the proposed solution:

- Processing Subsystem: An Espressif ESP32 microcontroller was selected since it provides a good tradeoff among embedded computing power, low cost and energy efficiency [89], as well as abundant General Purpose Inputs Outputs (GPIOs), built-in Wi-Fi and Bluetooth communications, and an I2C interface for communicating with the sensor interface.
- Thermal Imaging Sensor: A FLIR Lepton 3.5 was selected. It is a small (11.8 x 12.7 x 7.2 mm) but very capable sensor, which provides a very good trade-off between performance and price. Specifically, the sensor uses an Uncooled VOx microbolometer for thermal imaging detection, and is capable of outputting 160 x 120 pixels resolution images with a 57º horizontal FOV, it has a thermal sensitivity of less than 50 mK (0.050 ºC) and features a power dissipation of 150 mW while operating (650 mW during shutter events and 5 mW in standby).
- Sensor Interfacing: Allows for integrating the processing subsystem with the selected sensor. For such a purpose, a PureThermal2 was selected, which is a hackable Smart I/O Module. It has a built-in Micro-USB port and supports communications via I2C, UART and JTAG. It also includes a STM32F412 microprocessor that allows for executing on-board image processing. All this is provided in a compact form factor of only 30 × 18 mm.

4.3.2. Implemented Software

The developed software can be divided into three different parts: (1) Thermal Images Dataset; (2) Image Processing Module; and (3) adopted Human-detection Methods:

1. **Thermal Images Dataset**: To evaluate and validate the developed human-detection methods, a new thermal image dataset was obtained. For such a purpose, a Raspberry Pi 3B was used together with the PureThermal2 and the FLIR Lepton 3.5. A script was created to capture an image once every minute. During the acquisition phase, 1114 images were captured and added to the dataset.
2. **Image Processing Module**: Every image frame needs to be processed before applying the developed human-detection methods in order to uniformize the dataset. Specifically, every captured frame results in a 3D matrix of $160 \times 120 \times 3$, corresponding to width, height, and color space (RGB), respectively. Such a matrix is first processed to convert it to gray scale, removing the third dimension of the matrix, thus resulting in a $160 \times 120$ pixels images, where every pixel corresponds to its intensity. Then, with the frame in gray scale, the image is smoothened by applying an isotropic Gaussian smoothing. Previous steps are illustrated in Figure 6.
3. **Human-detection Methods**: This is arguably the most critical part of the developed software, since it is responsible for accurately detecting the presence of humans within a given frame. For this software development, a hybrid detection approach was devised around one simple principle: every human is considered a body of heat, so, any body of heat with a shape resembling a human should be considered as one. This principle is only valid when using thermal imaging cameras, since traditional RGB cameras are not capable of detecting sources of heat. The devised hybrid detection mechanism consists of two methods: one capable of detecting movement and a second capable of detecting regions of interest within a frame. Both methods run in parallel, so, if one of them determines that there is a human present, the result is predicted as a positive detection. Therefore, a negative detection will only occur if both methods determine that there is no human present. Figure 7 depicts the proposed detection approach, using the terms "Method A" and "Method B" for the two developed approaches.

    - **Method A** focuses on movement detection and follows the flowchart shown in Figure 8. As can be observed, it makes use of two frames: one that will serve as a starting point for the comparison, and another one that is compared to the previous. Thus, this method uses direct subtraction of image matrices to generate a new frame. This operation can yield two possible outcomes:

        (a) The result of absolute difference results in a new frame with pixel values averaging zero or near zero. This means that the new frame has no new significant updates; therefore, it can be concluded that no significant movement was detected. If this happens, the new frame will serve as the new background frame for the next comparison. This assures that the next comparison will always be performed with the most recent reference frames, so more accurate comparisons can be performed, and environmental factors can be compensated, such as temperature increases during the day and other external factors.

(b) The absolute difference results in one or more group of pixels (also called active pixels) that have values near reference values. This means that there was a major update in the frame, so the likelihood of existing movement is very high. To achieve higher confidence when detecting movement, the obtained result is compared with a threshold. Such a threshold was calculated by considering the number of pixels in the image. For a frame of 19,200 pixels (i.e., image resolution of 160 × 120 pixels), if at least 5% (i.e., 960 pixels) of those pixels are considered as active pixels, then it can be determined with certainty that there was a significant amount of movement.

- **Method B**, whose flowchart is depicted in Figure 9, uses a Region of Interest Approach that consists in using a single 160 × 120 area that is then divided into four 80 × 60 pixel quadrants. An example of these quadrants is shown in Figure 10. Such four quadrants, on their own, do not contain much information, but when compared with the complete frame, they can help to determine whether those quadrants have any significant value. This comparison is made by using the mean value of all the pixels in the total frame with each image quadrant, so, if the original frame (i.e., 19,200 pixels) has a mean value of X, a quadrant (i.e., 4800 pixels) will be considered of interest when its mean is at least 20% greater than X. This method is fast and simple, and presented good results, as will be shown later in Section 4.4. Moreover, it is worth pointing out that Method B shows increased accuracy in detecting static persons in different parts of the frame. However, its performance decreases when the body of heat is in the center of the frame, which results in dividing it into four different quadrants. Nonetheless, this method is easy to implement and customize, as its deployment can be targeted into resource-constrained devices. For instance, a single quadrant could be monitored or, considering the room layout shown in Figure 10, only two specific quadrants could be monitored.

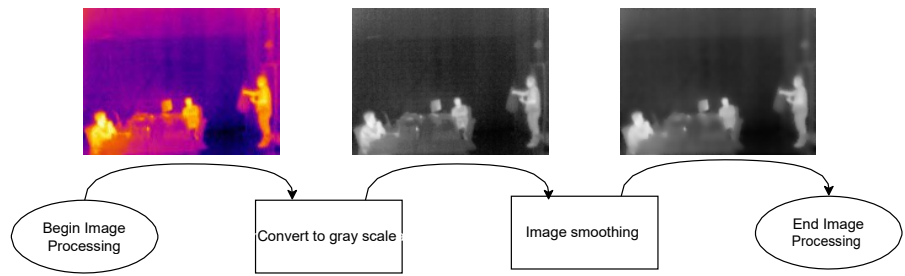

**Figure 6.** Steps performed for processing thermal images.

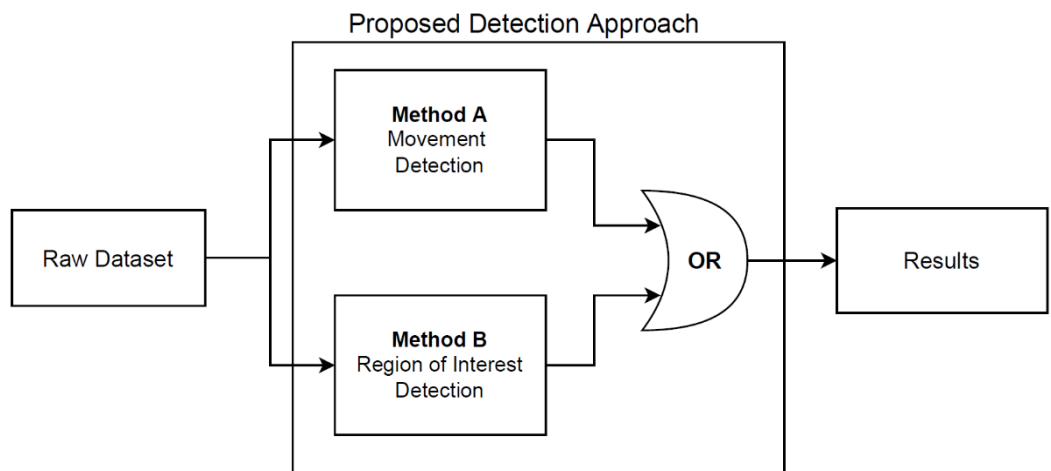

**Figure 7.** Proposed hybrid human-detection approach.

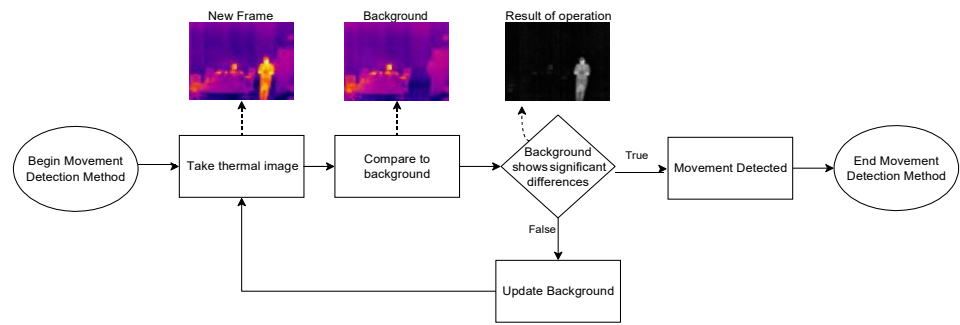

**Figure 8.** Flowchart for Method A (movement detection approach).

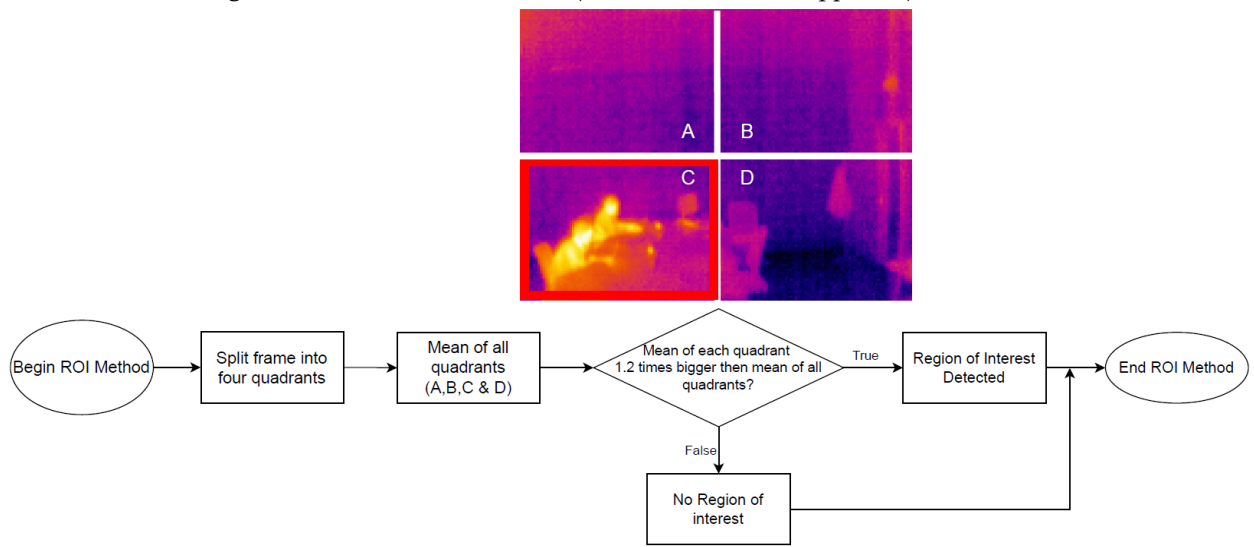

**Figure 9.** Flowchart for Method B (region of interest method).

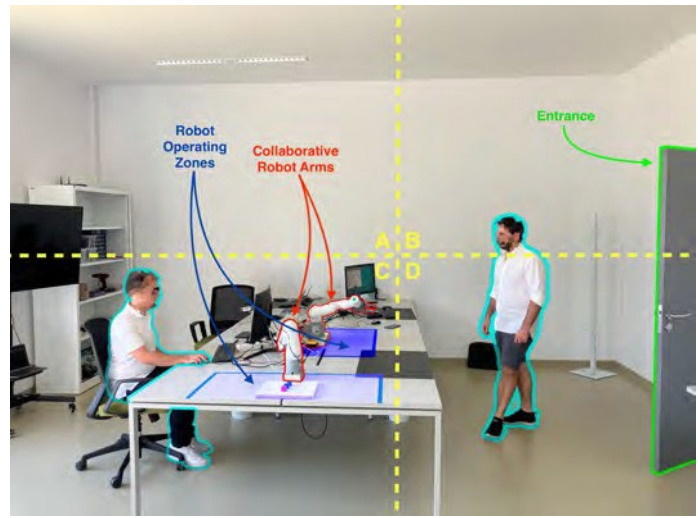

**Figure 10.** Setup of the experimental testbed in a research lab.

*4.4. Experiments*

4.4.1. Experimental Testbed

To evaluate the accuracy of the developed human-detection software in a real environment, an experimental testbed has been built. For such a testbed, to ease experiment automation and data collection, a Raspberry Pi 3B was used, which includes an ARM Quad Core 1.2 GHz Broadcom BCM2837 64bit SoC and 1 GB LPDDR2 RAM, together with the PureThermal 2 (which was attached through a Micro-USB connector) and the FLIR Lepton 3.5 sensor (as it is illustrated in Figure 11). The testbed was then placed on a vantage point overlooking a research lab (shown in Figure 10), which is frequently used by people and where collaborative robot arms operate.

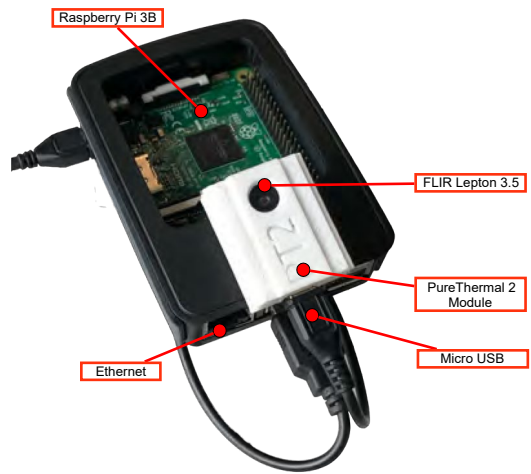

**Figure 11.** Hardware components of the mist node.

4.4.2. Performed Tests

For the performance evaluation tests, a script was first created to obtain the dataset to be used in the testing of the proposed human-detection methods. The script has been programmed to take four consecutive images every second and has been executed during four days. During that period, 1114 thermal images have been selected based on the manual identification of "human activity", to create the reference dataset used later to evaluate the proposed method. These 1114 thermal images have been manually annotated, where 1057 contained one or more persons, while the remaining 57 had no persons present on the frame. However, it is important to refer that, although in the mentioned 57 frames there were no persons present, there were different types of electrical and electronic equipment, generating and dissipating small amounts of heat. Then, the developed human-detection methods A and B were executed as previously illustrated in Figure 7.

4.4.3. Results

The obtained results are presented in this section as confusion matrices and together with a Key Performance Indicator (KPI) that measures the accuracy of the evaluated methods. Confusion matrices are often used to measure the performance of Machine Learning (ML) algorithms and their typical outputs results into two or more classes. In this case, the evaluated human detection only returns two output classes: positive (human detected) or negative (no humans were detected). The obtained result was then manually evaluated and categorized into four classes: True Positive (*TP*), True Negative (*TN*), False Positive (*FP*) and False Negative (*FN*). With these four classes, the accuracy was calculated according to Equation (1):

$$Accuracy = ((TP + TN)/(TP + TN + FP + FN)) * 100 \qquad (1)$$

The results of the individual application of method A are shown in the confusion matrix presented in Table 2. The results yielded 1101 positives and 12 negatives, of which 1057 were true positives, 44 were false positives and 12 were false negatives (there were no false negatives). It may seem at first sight that there is a higher than expected number of false positives, but this is because of the dataset in which the method was applied: some frames changed substantially from one day to another, so some new frames were flagged as positives despite being actually negatives. Nonetheless, considering a final Industry 5.0 implementation, having a higher number of false positives can be considered safer than having more false negatives. Overall, the accuracy of method A calculated through Equation (1), was 94.97%.

**Table 2.** Confusion matrix of results obtained by Method A.

|  |  |  | Ground truth (Manually Analyzed) | |
|---|---|---|---|---|
|  |  |  | 1069 | 44 |
|  |  |  | Positive | Negative |
| Predicted (Method A) | 1101 | Positive | 1057 TP (94.97%) | 44 FP (3.95%) |
|  | 12 | Negative | 12 FN (1.08 %) | 0 TN (0 %) |

The results of the individual application of method B are shown in the confusion matrix presented in Table 3. As it can be observed, 1027 frames were flagged as true positives, 11 as false positives, 48 as true negatives and 28 as false negatives. These results show that method B presents an accuracy of 96.5%. In comparison with method A, method B is not only more accurate, but it was also better at detecting static people and at dealing with sudden changes in the frame layout.

**Table 3.** Confusion matrix of results obtained by method B.

|  |  |  | Ground truth (Manually Analyzed) | |
|---|---|---|---|---|
|  |  |  | 1055 | 59 |
|  |  |  | Positive | Negative |
| Predicted (Method B) | 1038 | Positive | 1027 TP (92.19%) | 11 FP (0.99%) |
|  | 76 | Negative | 28 FN (2.51%) | 48 TN (4.31%) |

Finally, Table 4 shows the results obtained when utilizing the hybrid approach. In such a case, both methods were used simultaneously, first applying method B and then method A. Both method A and B used the same parameters as for their individual evaluation. The obtained results indicate that there were 1040 true positive, 41 true negatives, 16 false positives and 17 false negatives, resulting in an overall accuracy of 97.04%. Therefore, the proposed hybrid approach requires more computational resources due to the simultaneous use of both detection methods, but it increases the overall accuracy of the system.

**Table 4.** Confusion matrix of the results obtained by the hybrid approach.

|  |  |  | Ground truth (Manually Analyzed) | |
|---|---|---|---|---|
|  |  |  | 1057 | 57 |
|  |  |  | Positive | Negative |
| Predicted (Both Methods) | 1056 | Positive | 1040 TP (93.35%) | 16 FP (1.44%) |
|  | 58 | Negative | 17 FN (1.53%) | 41 TN (3.68%) |

4.4.4. Result Analysis

The proposed hybrid approach allows for a good trade-off between performance and computing power, because it does not rely on compute-intense algorithms to identify humans or shapes of humans in a frame like most ML or AI algorithms currently available. Moreover, both methods A and B only use basic matrix operations to perform the analysis and comparison, and no previous training is required. Thus, the proposed hybrid solution can be easily implemented in an edge-like device (e.g., an ESP32) that has low energy consumption, low footprint, low complexity and low cost.

In case of trying to achieve a 99.9% accuracy, some trade-offs would have to be made, mainly in the computational capabilities of the proposed solution. For example, it can be compared with some algorithms such as YOLOv3, which is still considered to be a very accurate and fast state-of-the-art object detection algorithm, but which is only evaluated on powerful GPUs (Graphics Processing Unit). This means that the accuracy of YOLOv3 comes at a high computational cost. For instance, previous experiments with YOLOv3 show that even when using a powerful GPU like a GeForce GTX 1080 Ti, it takes the algorithm 200 ms to detect objects in a single image, while real-time detection requires this value to be lower than 40 ms [90]. With the new YOLOv4, implementations like the ones presented in [91] show high accuracy values (i.e., 99.28%) in detecting a variety of objects as well as persons, but the minimum hardware requirements are 8 GB RAM, 500 GB of hard disk, a 64-bit processor and a GPU. Another example is described in [92], where the authors make use of various AI algorithms to achieve an accuracy of 96.5% when using a traditional desktop computer, but propose the use of a Jetson Nano, which is cheaper than traditional desktop and laptop solutions, and is more powerful than the proposed ESP32, but it is still considerably more expensive (the Jetson Nano currently costs roughly 100 euros, while an ESP32 can be purchased for less than 10 euros).

For the sake of clarity, Table 5 compiles and compares the obtained results with the ones of the previously mentioned papers. It is worth mentioning that there are different factors that impact the performance of the compared systems, so a fair comparison should consider the resolution of the captured images (the higher the resolution, the longer it takes to process an image), the involved processing hardware (the more powerful the hardware, the faster it processes each image) and the used image dataset (performance varies depending on the specific set of processed images). Thus, taking the previous considerations into account, it can be observed that the proposed solutions provide a good trade-off among detection accuracy, maximum latency and IoT node cost.

**Table 5.** Performance comparison of the proposed solution with state-of-the-art systems.

| Reference | Type of System | Architecture | Main Hardware | Detection Algorithm | Accuracy | Maximum Detection Latency | Cost |
|---|---|---|---|---|---|---|---|
| [80] | CPHS for preserving safety when working with cobots | Cloud Computing | Each IoT node has an 8 Megapixel camera and several, an ultrasound sensor, a stepper motor and a Raspberry Pi. Images are preprocessed on the Raspberry Pi and sent to a remote Linux server that carries out detection through its GPU. | HoG<br>Viola Jones<br>YOLO v3 | 70%<br>77%<br>99% | 201 ms<br>191 ms<br>151 ms | Medium (IoT nodes cost less than 100€, but a powerful server is required) |
| | Algorithm | | Powerful GPU (e.g., a webcam. | | 99% | 200 ms | Medium (IoT nodes cost less than 50€, but a powerful server is required) |
| [91] | Algorithm implementation | n/a | Webcam and PC with at least 8 GB of RAM, 500 GB of disk drive, 64-bit CPU and a dedicated GPU. | YOLO v4 | 99.28% | n/a | Medium |
| [92] | CPHS for measuring social distancing | Cloud Computing | 8 Megapixel webcam and an NVIDIA Jetson Nano (ARM57, 1.43 GHz, 64-bit CPU, 4 GB of RAM, 128-core GPU). | Based on YOLO v2 | 95.60% | 37 ms | Low (less than 200€ per IoT node) |
| This Work | CPHS for avoiding industrial operator exposition to dangerous situations | Edge/Mist Computing | Low cost SBC (Raspberry Pi 3B), thermal imaging sensor, PureThermal2 adapter. | Method A<br>Method B<br>Hybrid Method | 94.97%<br>96.50%<br>97.04% | 7 ms<br>6 ms<br>10 ms | Low (roughly 200€ per IoT node) |

## 5. Discussion

*5.1. Key Findings*

The design, implementation, and practical evaluation of a CPHS for Industry 5.0 applications allowed us to collect relevant findings based on hands-on experience that is worth summarizing for guiding Industry 5.0 researchers and engineers:

- After an analysis of the state-of-the art of the recent CPS and CPHS available in literature, surprisingly, it was found that most of current works only comply with one or two of the Industry 5.0 principles (human-centricity, sustainability, and resilience). In addition, most of the current CPHSs are mainly used in healthcare. Furthermore, there are not many cost-effective systems with low computational complexity to detect human presence.
- In addition, it is worth noting that the proposed novel hybrid mist and edge communications architecture, the Industry 4.0/5.0 enabling technologies discussed throughout this article were considered having a hostile industrial scenario in mind with real-time or near real-time and intensive computing requirements.
- Managing Quality of Service (QoS) in CPHSs communications is an important issue, since the implementation of CPHSs usually depends on state-of-the-art communications like 5G. QoS analysis is an important topic, since CPHSs are frequently used in industrial applications where latency requirements are strict. For example, 5G technologies present built-in mechanisms for defining, implementing, controlling, policing, and monitoring QoS.
- The proposed IoT Thermal Imaging Safety System is a cost-effective solution designed to trigger different safety states in manufacturing processes that rely on collaborative robotics. It enables the protection against possible incidents or accidents or the optimization of the energy efficiency of the deployed industrial devices and machinery. However, the proposed system should be used, in parallel, with a secondary safety redundant system, because although the system performance is high, detection accuracy is still not 100%.
- The system does not use centralized ML or AI algorithms that require prior training. It demands low-energy consumption, low footprint, low complexity, being a cost-effective solution that can be easily deployed into IoT architectures, with minimal intervention and tuning requirements, thus being able to operate in different environmental conditions.
- The adoption of Safety by Design (SbD) practices in CPHSs has a relevant impact in reducing accidents, by predicting, recognizing, and avoiding hazards, rather than retrofitting CPHSs after an accident has occurred. In this case, user safety is prioritized during the design and development of CPHSs.
- The design and development of CPHSs rely on distributed autonomous computing devices that require Distributed Real-Time Operative Systems (DRTOS) capabilities to perform inter-dependent critical tasks, such as motion control, computer vision, and machinery actuation, which demand deterministic execution and low-latency networks.

*5.2. Challenges*

The CPHS to be developed should be first analyzed in terms of requirements and communications architecture, while keeping in mind Industry 5.0 principles. In addition, hardware should be chosen according to the selected communications architecture and the specific application.

After all the performed analyses, it can be stated that Industry 5.0 still faces many challenges, specially in human–machine collaboration. These obstacles have to be overcome by researchers and developers who will have to cope with the additional issues presented by smart manufacturing environments as well as the unique nature of their operations and networks.

Manufacturers have long placed a high focus on workplace safety. Many of them, with the aim to introduce collaborative robotics in the manufacturing processes, are reeval-uating their workplace safety procedures on the shop floor, although, redundant safety systems are still a standard and are required by legislation in force. For example, (1) monitoring a collaborative robotic arm with a camera; (2) using a safety laser scanner for detecting objects in its surroundings; (3) a simple emergency stop button (the classic red button); or a combination of them, must be considered if a risk assessment procedure identifies an "unacceptable risk of injury", due to an unexpected collision with a human body when the robot is operating at higher speeds. A relevant challenge would be the reduction of the redundant safety dimension. However, this approach would result in the design of a more accurate detection approach, in which, AI-based solutions may play an important role. Therefore, it is more important than ever for manufacturers to make use of the advantages of cutting-edge AI technologies, such as Edge Intelligence, or Edge-AI, not only for increasing workplace safety, but also to reduce the redundant safety dimensions. Thus, future work should include the inclusion of an Edge-AI approach that complies with the three Industry 5.0 principles (Human-centricity, Sustainability and Resilience) and a procedure to compare its performance with the proposed solution. The use of Edge-AI reduces bandwidth in communications and enhances prognosis and decision-making through the use of device prediction models, which allows the execution of ML algorithms directly on the embedded device while taking advantage of a near-perfect accuracy.

*5.3. Relevant Guidelines for Future Researchers*

In order to guide future developers, this section describes the main stages of the development of future application cases (a summary of such stages is depicted in Figure 12):

1. Problem statement. it is first necessary to analyze the stakeholders' needs, to define the application scenario and to determine the involved processes.
2. Review of the state of the art. Relevant works on CPHSs have to be analyzed in terms of compliance with the principles of Industry 5.0. In addition, a detailed analysis of available technologies and communication architectures must be performed.
3. Application case analysis. The first step is to describe the selected use case in general terms, with emphasis on its main objectives. Specific operational and technical requirements must be detailed and analyzed along with the application scenario itself.
4. Design. Industry 5.0 principles (human-centricity, sustainability, and resilience) have to be taken into account throughout all the design process. An adequate trade-off must be obtained between the following aspects:
   - Human-centricity: The system must be designed with the human needs in mind (e.g., human–machine collaborative scenarios) in a context of no full automation. Aspects like upskilling and reskilling of operators or technical assistance have to be carefully considered, adapting the production process to the needs of the worker and using technology to guide and/or train humans.
   - Sustainability: The system must be designed with operational and resource efficiency at the core. In terms of hardware, low-cost and low power hardware must be considered. The communication architecture must be designed to minimize energy consumption. In terms of software, service-orientation and personalization have to be considered. In addition, accountability and transparency, data sharing (e.g., vertically or horizontally communications (in real-time when needed)) must be included [31]. The designed and developed services must be environmentally friendly across their lifetime and function within a circular economy framework.
   - Resiliency: Real-time or near real-time communications must be provided together with resilience to cyberattacks (e.g., redundancy) and privacy measures.
5. Implementation. The designed CPHS will be implemented using the selected hardware and software.
6. Experimental validation. The developed system needs to be tested in the laboratory. After an analysis of the obtained results, the system must be evaluated in terms of performance with state-of-the-art systems.

7. Results report. The key findings based on practical experience may be reported to guide future researchers and developers.

The following stages, which are out of the scope of this article, should include the validation in real-world scenarios to determine whether the Industry 5.0 CPHS meets the requirements established in the analysis stage, the integration with other functional services (e.g., information management, third-party systems), the validation by the operators, and the transition to operation (e.g., OT&IT convergence) and maintenance phases.

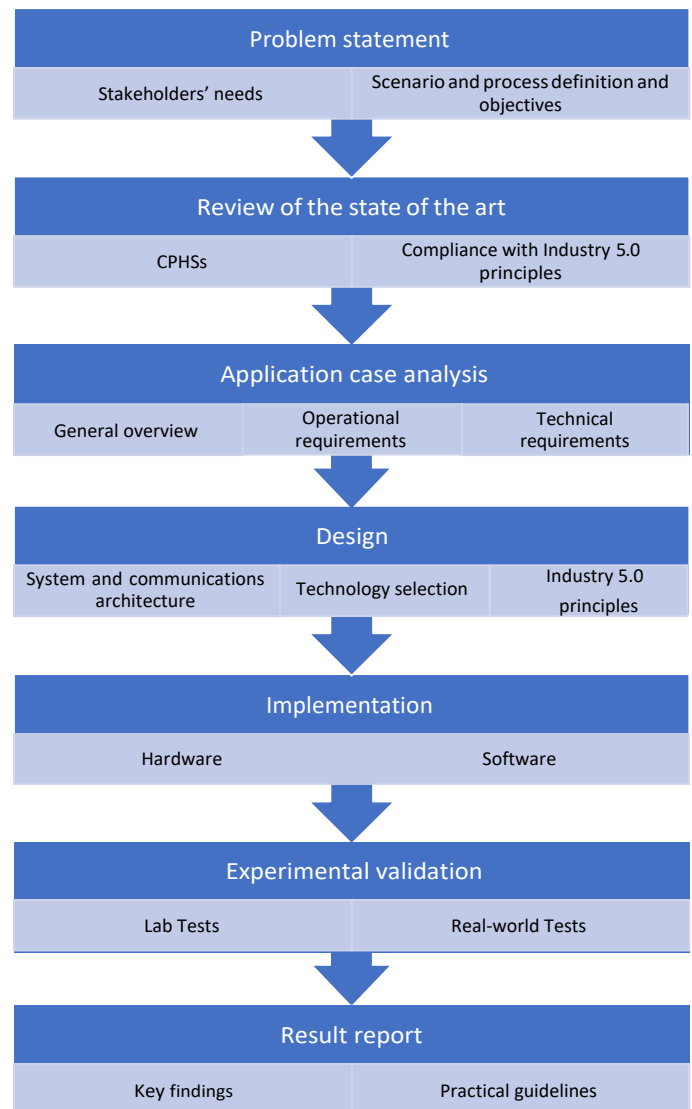

**Figure 12.** Proposed Industry 5.0 CPHS development methodology.

## 6. Conclusions

This article presented a CPHS for Industry 5.0 applications that makes use of low-cost hardware to detect human proximity and thus avoid risky situations in industrial scenarios. The proposed CPHS was evaluated in a real-world scenario with the main goal of increasing human safety within manufacturing processes that rely on collaborative robots or other machinery. The chosen strategy makes use of a hybrid edge computing architecture that is both affordable and efficient and takes advantage of smart mist computing nodes to analyze thermal images and to warn about industrial safety hazards. After a thorough review of the state of the art, the design and implementation of the proposed edge/mist computing CPHS were described, and its performance evaluation was detailed. The obtained results show that the three devised human-detection algorithms reach a detection accuracy of up to 97.04% with a maximum latency of 10 ms, while having a low computational complexity that allows for implementing them in resource-constrained devices like SBCs or SoCs. As a consequence, the solution provides a good trade-off between cost, accuracy, resilience, and computational efficiency.

Altogether, this article not only presents a novel edge/mist computing CPHS, but also provides useful guidelines to practitioners, scholars, and industry frontline managers, that may be useful to develop the next generation of Industry 5.0 CPHSs.

**Author Contributions:** conceptualization, P.F.-L., S.I.L. and T.M.F.-C.; methodology, P.F.-L., S.I.L. and T.M.F.-C.; investigation, P.F.-L., D.B. and T.M.F.-C.; writing—original draft preparation, P.F.-L., D.B., S.I.L. and T.M.F.-C.; writing—review and editing, P.F.-L., S.I.L. and T.M.F.-C.; supervision, T.M.F.-C.; project administration, T.M.F.-C.; funding acquisition, P.F.-L., S.I.L. and T.M.F.-C. All authors read and agreed to the published version of the manuscript.

**Funding:** This work has been performed under the scope of project TECH - Technology, Environment, Creativity and Health, Norte-01-0145-FEDER-000043, supported by Norte Portugal Regional Operational Program (NORTE 2020), under the PORTUGAL 2020 Partnership Agreement, through the European Regional Development Fund (ERDF). P.F.-L. and T.M.F.-C. have been supported through funds ED431G 2019/01 provided by Centro de Investigación de Galicia "CITIC" for a three-month research stay in Instituto Técnico de Viana do Castelo between 15 June and 15 September 2022. This work has also been funded by the Xunta de Galicia (by grant ED431C 2020/15), and by grant PID2020-118857RA (ORBALLO) funded by MCIN/AEI/10.13039/501100011033.

**Acknowledgments:** P.F.-L. and T.M.F.-C. would like to thank CITIC for its support for the research stay that led to this article. CITIC, as Research Center accredited by Galician University System, is funded by "Consellería de Cultura, Educación e Universidades from Xunta de Galicia", supported in an 80% through ERDF Funds, ERDF Operational Programme Galicia 2014-2020, and the remaining 20% by "Secretaría Xeral de Universidades" (Grant ED431G 2019/01).

**Conflicts of Interest:** The authors declare no conflict of interest.